# VIS-iTrack: Visual Intention through Gaze Tracking using Low-Cost Webcam

Determining Human Intention for Perceiving Textual or Graphical Information from Eye Gaze Information

Shahed Anzarus Sabab[a, b, c, d, e], Mohammad Ridwan Kabir[a, b, c, 1], Sayed Rizban Hussain[a, b, c], Hasan Mahmud[a, b, c], Md. Kamrul Hasan[a, b, c], Husne Ara Rubaiyeat[f,]

[a] *Systems and Software Lab (SSL)*
[b] *Department of Computer Science and Engineering*
[c]*Islamic University of Technology (IUT), Gazipur, Bangladesh.*
[d]*Department of Computer Science*
[e] *University of Manitoba, Winnipeg, Canada*
[f] *National University, Bangladesh.*
sababsa@myumanitoba.ca, {ridwankabir, rizban, hasan, hasank}@iut-dhaka.edu; rubaiyeat@yahoo.com.

Human intention is an internal, mental characterization for acquiring desired information. From interactive interfaces containing either *textual* or *graphical* information, intention to perceive desired information is subjective and strongly connected with eye gaze. In this work, we determine such intention by analyzing real-time eye gaze data with a low-cost regular webcam. We extracted unique features (e.g., *Fixation Count, Eye Movement Ratio*) from the eye gaze data of 31 participants to generate a dataset containing 124 samples of visual intention for perceiving *textual* or *graphical* information, labeled as either *TEXT* or *IMAGE*, having 48.39% and 51.61% distribution, respectively. Using this dataset, we analyzed 5 classifiers, including *Support Vector Machine* (*SVM*) (*Accuracy*: 92.19%). Using the trained *SVM*, we investigated the variation of visual intention among 30 participants, distributed in 3 age groups, and found out that young users were more leaned towards *graphical* contents whereas older adults felt more interested in *textual* ones. This finding suggests that real-time eye gaze data can be a potential source of identifying visual intention, analyzing which intention aware interactive interfaces can be designed and developed to facilitate human cognition.

Keywords: Human-Computer Interaction, Visual Intention Detection, Eye Gaze, Kalman Filtering, Saccades, Fixation, Support Vector Machine, Intention Aware Interfaces.

## 1 INTRODUCTION

Human cognitive processes, such as thinking, learning, remembrance, decision-making comprise a significant part of Human-Computer Interaction (HCI). These processes lead to intentions that are implicit in nature and cannot be easily interpreted [1]. So far, researchers have tried to determine such intentions in different scenarios of HCI [1]–[6] using eye gaze data. The eye movement patterns vary across different implicit intentions. The *saccadic* and *fixation* movements of the human eyes are vital to the determination of user intention [7]. During the *saccadic* movement, both of our eyes move simultaneously and quickly between two points in the visual field without collecting much useful information [7], [8]. However, during *fixation*, occurring between two *saccadic* movements, the human eyes tend to focus on a certain *Region of Interest* (*ROI*) in the corresponding visual field, typically for a period of 200-600ms [7]–[9] known as the *fixation duration*, where perception occurs.

Information on interactive interfaces is usually presented in two modes, either *textually* or *graphically*, on a computer screen. The intention to retrieve *textual* or *graphical* information from such interfaces is a human cognitive process that is implicit in nature [8]. Therefore, analyzing the *saccadic* and *fixation* movements of the human eyes while interacting with such interfaces, recorded with different eye-tracking technologies, can create a pathway for determining the visual intention of a user to retrieve information from a particular type of content. The human intention varies within different groups i.e., demographics, gender [10], and age [8]. Due to this subjective behavior, the *User Interface* (*UI*) or *User Experience* (*UX*) designers can present relevant information based on different types of applications targeted towards different user groups. Therefore, analysis of users' gaze information through features such as *eye movement time*, *fixation time*, and *jerky movement time* is subjected to be a viable option for detecting users' intention and their goal formulation process [3], [11]–[14].

In this work, we present a system to detect and classify human intentions to perceive *textual* or *graphical* information from interactive interfaces through classical *Machine Learning* (*ML*) approach based on real-time eye movement tracking using a low-cost general-purpose webcam. Leveraging the eye-tracking strategies using such webcams, reported in studies

---
[1] *Corresponding Author*

[15]–[18], we developed our tracking system using the webcam feed as a reference. In our approach, we have extracted the pupil coordinates of the user's eyes while they are focusing on a particular section of the screen. We have improved the eye detection by adding heuristic calculation [19] of the relative eye regions and only considered small region (where eyes can reside) to remove noises. Furthermore, to increase classification accuracy, we have smoothened the *pupil movement path* using the Kalman filter [20].

Our aim is to perform a case study where we extract key features from user's gaze data and analyze its impact on the *binary classification* of users' focus on either *textual* or *graphical* information presented on a computer screen. Therefore, we have explored 5 classifiers such as *K- Nearest Neighbors* (*KNN*), *Gaussian Naïve Bayes* (*GNB*), *Logistic Regression* (*LR*), *Support Vectors Machine* (*SVM*), and *Random Forest* (*RF*) to get an understanding of the type of classifier that is reliable for relevant studies.

Since we have processed real-time video feed from the webcam at *30 frames per second* (*fps*), each frame needs to be processed individually for tracking the user's gaze. Therefore, to ensure real-time eye gaze detection, we need to reduce the computational cost while maximizing the performance of the classifiers. This can be achieved with a minimalistic feature vector containing only the vital features. Therefore, we have defined a *feature vector* containing 8 features (4 features unique to each eye), using which we have generated a dataset to train and test the classifiers through *Repeated K-Stratified Fold Cross-Validation* [21] with K=10 fold and 5 repetitions. The classifiers are evaluated using different metrics such as the *Accuracy*, *Area Under the ROC Curve* (*AUC*), *Precision*, *Recall*, and *F1-score*.

As the movement of our eyes are coordinated, we cannot move them individually, in different directions. Again, due to binocular vision, we cannot perceive information from two different visual stimuli, one placed in front of each eye, simultaneously [22]. Therefore, while focusing on a visual stimulus, the eye movement paths vary for both eyes. We investigated the performance of the classifier having features from a single eye (left) vs both eyes. Our result suggests that the performance is better while using features from both eyes (details in Section 3.7).

Furthermore, to showcase one application of our technique, we conducted a user study where we determined the type of information that was focused on more (*textual* or *graphical*) across different age groups using *SVM* as a trained classifier. We have adopted SVM due to different factors such as – 1) Having the highest performance while trained with the features from both eyes (*Accuracy:* 92.19%), 2) Faster inferencing time (i.e., avg inference time = 0.8 milliseconds/sample), 3) With a comparatively smaller dataset, SVM is less vulnerable to overfitting, favoring generalization [23].A brief overview of our proposed approach of determining visual intention is shown in Fig. 1.

To summarize our contributions: 1) We develop a feature extraction system, which takes real-time gaze information for feature engineering. 2) We verify that a machine learning model leveraging the gaze features can be used in classifying visual intention (TEXT vs IMAGE), and 3) We demonstrate one application of visual intention to present adaptive user-interface to users of different age groups.

In the next section, we present the literature review followed by an explanation of the proposed approach of determining the visual intention of a user to retrieve *textual* or *graphical* information from interactive interfaces. We then elaborate on the user study and analyze the outcomes of the study. Finally, we summarize our observations and give a direction on future works.

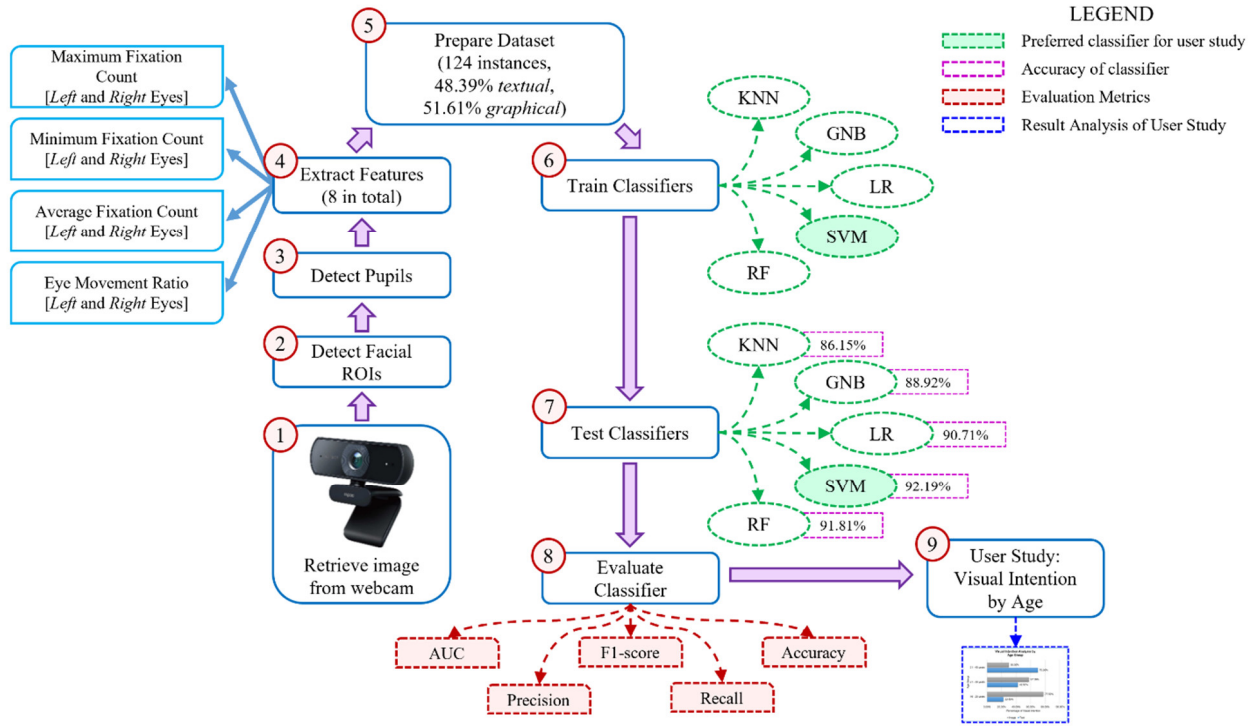

Figure 1: Overview of the proposed approach of determining visual intention for *textual* or *graphical* contents in interactive interfaces.

## 2 RELATED WORKS

The human eyes are sources of rich visual information and can provide an insight of the human intention through analysis of their eye gaze data. This is an emerging field of research where several studies have been conducted to date.

Pupil detection and tracking is a vital and the most difficult step for eye gaze tracking [15]. In order to accomplish this, researchers in this domain have used eye-trackers such as RED Eye-tracker [24], ASL 504 [8], Tobii [1], [2] or other head-mounted devices for extracting different features such as *pupil size variation*, *eye scan path*, *saccades*, *fixation*, and *ROIs* [1], [2]. However, these devices are costly, making such systems infeasible and inaccessible to the general mass. To resolve these shortcomings, studies [15]–[18] have analyzed the feasibility of tracking eye gaze using low-cost regular webcams such as Logitech C300 [15], Logitech Quick Cam Pro 9000 [17] in real-time with reasonable accuracy, resulting in a cost-efficient solution and eliminating the burden of a head-mounted device. Using *Circular Hough Transform* (*CHT*) [25], [26], eye pupils have been tracked using such webcams after processing the camera feed in [15]. Researchers have also explored neural networks for eye gaze tracking due to its robustness to noisy data [16]. A comparative analysis of different algorithms such as *Cumulative Distribution Function* (*CDF*), *Projection Function* (*PF*), *Edge Analysis* (*EA*) have also been carried out to understand the type of pupil detection approach best suited for regular webcams [18]. Motivated by the previous works, we chose to work with a low-cost regular webcam for eye gaze tracking, providing a cost-effective solution to visual intention detection.

Researchers in [1] have attempted to classify implicit human intention as either *navigational* or *informational* in both indoor and outdoor environments. They have extracted key features of the eye such as *fixation length*, *fixation count*, and *pupil size variation* as reliable features from eye gaze data, recorded using the Tobii 1750 eye-tracker. They have used two types of classifiers: (1) *Nearest Neighborhood* (*NN*) and (2) *SVM*. The average accuracy of the proposed system considering these two classifiers was over 85%. The major drawback of their recommended architecture is the dependency on an expensive eye-tracking system. Apart from identifying human intention as *navigational* or *informational*, a new type of intention, *transactional*, has also been explored in case of web searching [4] through analysis of a web search engine log containing about a million and a half queries from numerous hundred thousand users. The authors have developed an automated three level classifier with features such as *Color Histogram, Color Spread,* and so on. The reported accuracy of their approach was 74%. However, there are a few limitations of this study, as stated by the authors themselves. First, the user intent for a particular query was annotated to single search intent (*informational, navigational,* or *transactional*), manually from the search engine log whereas a particular search may have multiple possible intention due to the implicit and hard to interpret nature of human intentions. Second, there is an inherent shortcoming of relying solely on data from transaction logs, which involves not having access to the users for correctly identifying their intent. The accuracy of their classifier is valid given that the manual classification of queries is correct. However, the stated accuracy of their classifier over such a large dataset is an indication of its robustness. In other research works [5], having similar classification objectives as [1] and [4], user queries were manually classified from a relatively smaller transaction log for gaining an insight into the proportions of various types of search intents of the users. Findings of these studies suggest that about 40% of such intentions were *informational*. However, one limitation of [5] is that it was not verifiable whether the manual classifications of intents were in fact, the original intent of the user. Considering these research works, we have considered fixations and eye movement patterns as relevant features from gaze information.

With the motivation to facilitate real-time HCI through analysis of complex human behavior, a framework that automatically detects visual attention to task-related objects or *ROIs* has also been proposed [27], using pre-trained *ML* models for object detection and classification. Recently, eye gaze tracking in determining visual intention in Human-Robot Interaction (HRI) has gained attention from researchers. GazeEMD, an approach to detect visual intention of users [28], facilitating user interaction with a robotic manipulator in picking up an object with eye gaze has been explored as well. Extending the application of eye gaze in wearables in real-time, *fixation points* have been used to determine users' intention to know about an object in their visual field [29]. Having determined users' intention, only the intended object was recognized. It is interesting to note that all of these were done in real-time, resulting in a more immersive experience. Prior studies have also explored the application of viewers' gaze to discover similarities between the visual intention of users in comic art [24]. It is evident from these research works that human eye gaze data is a reliable means of determining visual intention in a lot of intriguing application areas. Therefore, we have showcased one of the many possible applications of visual intention detection using our proposed approach.

In the next section, we discuss our proposed approach, where we try to determine the visual intention of a user to retrieve *textual* or *graphical* content from interactive interfaces.

## 3 PROPOSED APPROACH

In this section, we elaborate on our proposed approach of processing the webcam feed for *face detection* and outlining the eye regions. Once the left and the right eye regions are detected, they are processed for *iris and pupil detection,* followed by *pupil tracking*. Next, we discuss feature extraction, dataset generation and analysis of feature properties. Finally, we present our approach of classifying visual intention followed by the evaluation of classifier performance.

### 3.1 Webcam Image Processing

Our motivation behind using a generic webcam (*Logitech C920*) is to develop a cost-efficient solution for determining visual intention using users' gaze. One of the challenges of this study is to track the movement of the eye pupils using the low-resolution *Real-Time Video Feed* (*RTVF*) of the webcam [15]. For our purpose, we have used an image resolution of 800×600 at 30 fps. For pupil tracking, we have followed a six-step process. The steps are: (1) Regions containing the viewer's face (Fig. 2a, *Legend:* "*Facial Region*") is detected in real-time using the *Viola-Jones* [30]–[33] algorithm, facilitating rapid face detection, which is an essential aspect of our proposed approach. We have used a pre-trained Haar-Cascade face detection model based on the sample dataset [34], containing thousands of negative and positive facial image information. At the beginning of each user session a calibration phase is required where the *RTVF* is processed for manual tuning of image-related parameters (min-max object size, min neighbors, etc.) [35] until viewer's face is detected. The calibration is essential because of how light reflects differently for people with different skin tones. (2) The probable eye regions (Fig. 2a, *Legend:* "*Eye Regions*"), are identified using a modified heuristic calculation based on [19]. (3) The *left* and the *right* eyes are detected from the respective eye regions (Fig. 2a, *Legend:* "*Detected Eye Regions*") in the same approach as face detection, using two separate datasets [36]. Since these eye regions are smaller than that of the face (Fig. 2a, *Legend:* "*Facial Region*"), feature matching in this reduced space enhances the accuracy of eye detection. (4) The detected eyes are further processed to enhance the contrast using Histogram Equalization [37], followed by inversion and binary thresholding, and finally, transformation using *Gaussian Pyramid* [38]. For instance, Fig. 2b shows the processed image of the right eye. (5) From the processed image of each eye, the *iris* region is detected using *CHT* [25], [26]*,* and the *pupil* is considered as the center of this region. For example, the detected *iris* region and the *pupil* are outlined for the right eye in Fig. 2c. (6) Although the pupils are calculated as the center of the iris, at 30 fps, misdetections were observed in some frames. Potential reasons for such misdetections could be the occlusion of the iris and the pupils during eye blinks. This situation is undesired; therefore, we have introduced *Kalman* filter [20] in the pipeline, drastically reducing such cases. The coordinates of the pupils are then recorded for tracking their movements.

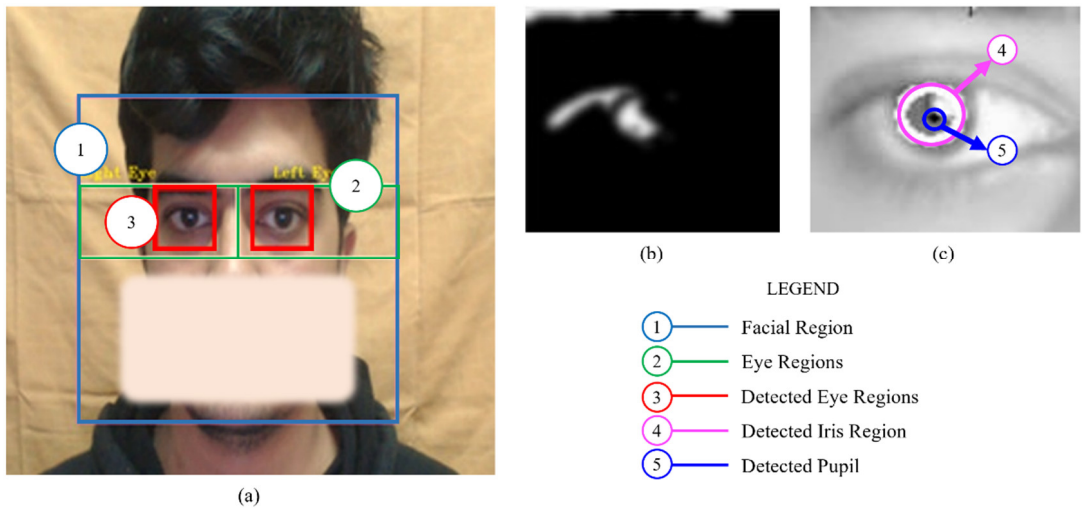

Figure 2: (a) *Facial* and *Eye Regions* of a user. (b) Detected and processed *Right* eye for iris and pupil detection. (c) Detected *Iris Region* and *Pupil* of the *Right* eye.

### 3.2 Feature Extraction

Eye gaze and its variations are mere considerations of the pupil movement and its variation, respectively. Distinguishable characteristics of pupil movement are noted from various contexts, which help in the determination of users' intention. After the pupils of both eyes have been tracked down, the next step is to extract unique features for classifying an intention to be either focused on *textual* or *graphical* contents. An important point to note is that since we are detecting human intentions using real-time tracking of viewer's gaze using a low-cost webcam, we focused on features that may be sufficient for accurately classifying visual intention with reduced computational complexity. Therefore, we have calculated 4 features such as *Maximum Fixation Count* (*MAX_FC*), *Minimum Fixation Count* (*MIN_FC*), *Average Fixation Count* (*AVG_FC*), and *Movement Ratio* (*MR*), for both *Left* (*L*) and *Right* (*R*) eyes, resulting in a total of 8 features for training the classifiers.

#### 3.2.1 Fixation Count

Gaze points detected using an eye-tracking system give an idea of a viewer's interest. When a collection of gaze points is very close to each other, a viewer is said to have fixated his/her focus on that *ROI*, containing the collection of gaze points. *Fixation duration*, defined as the amount of time spent on a particular *ROI*, is considered to be one of the most useful features for determining visual intention in the area of eye-tracking research [39], [40]. *Fixation Count* (*FC*) is defined as the number of gaze points in a particular *ROI*. The previous studies [41], [42] have shown that the *fixation duration* is longer when a user is looking at *graphical* content compared to *textual* ones. *Graphical* information includes *discrete* points of focus, whereas; *textual* information includes *continuous* points of focus. Based on prior studies [8], [12], [27], [28], [43], we considered *fixation duration* of 200ms as a threshold to define a fixation event. Intuitively, the *fixation counts* also increase proportionally with *fixation duration*. Based on these observations, in our study, we have calculated the *Minimum, Maximum,* and *Average Fixation Counts* for each of the eyes, resulting in 6 features for training the classifiers.

#### 3.2.2 Eye Movement Ratio

Apart from *fixation counts*, another distinguishable feature for intention detection is the *Movement Ratio* (*MR*) that depends on the *horizontal* and the *vertical* scan counts of an eye, as in (1).

$$Movement\ Ratio = \frac{Count\ of\ Horizontal\ Scans}{Count\ of\ Vertical\ Scans} \quad (1)$$

This feature is vital for determining such visual intentions, because, when the visual stimulus is *textual*, we observed that the horizontal movement of the eye is greater than its vertical movement [12], [43] whereas, for *graphical* information, both of these movements are of similar proportions [24]. Therefore, this ratio appears to be higher for *textual* information than that for *graphical* information. In our study, we have considered the *MR* of both eyes, resulting in 2 more and a total of 8 features for training the classifiers.

### 3.3 Dataset Generation

For preparing the dataset, 31 participants (Mean: 28.32 years, SD: 10.30 years, Male: 66.67%, Female: 33.33%) were recruited (i.e., via social media and word of mouth) for going through multiple images of *textual* and *graphical* contents. We considered a collection of 20 contents, consisting of 10 textual and 10 graphical ones. The word count in the *textual* contents ranged between 200 and 500, and the *graphical* contents involved images with several highlights. As, for our investigation, we chose to work with contents having an aspect ratio 4:3 and 16:9 because of the widespread usage in the media. From each collection of contents, each participant was asked to go through 2 randomly chosen unique *textual* and 2 randomly chosen unique *graphical* contents. Before initiating sample collection each participant was instructed about the task followed by a trial session which combinedly took roughly ~10 minutes. The trial session was introduced to give an essence of the real task and no data was recorded during this period. A pair of our sample contents used for dataset generation is shown in Fig. 3.

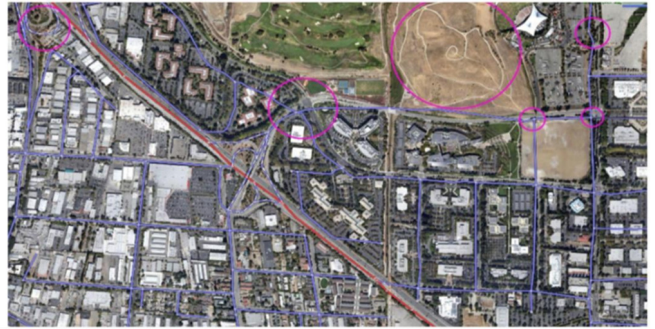

Figure 3: Sample images with *textual* (left) and *graphical* (right) contents used for dataset generation.

During the sample collection, each participant was given 5 minutes per content, while eye gaze data were recorded, analyzed, and the corresponding feature vector was generated. Therefore, for 4 contents each participant was given in total of 20 minutes. At the end of the experiment, the dataset, containing 124 feature vectors (*31 participants × 4 contents per participant*) as instances, was prepared. A workflow diagram of dataset generation is outlined in Fig. 4. Furthermore, since we used a regular webcam without infrared capability, sufficient lighting condition was ensured for proper data collection. A summary of the dataset is provided in Table I and an extract from the dataset is given in Table II.

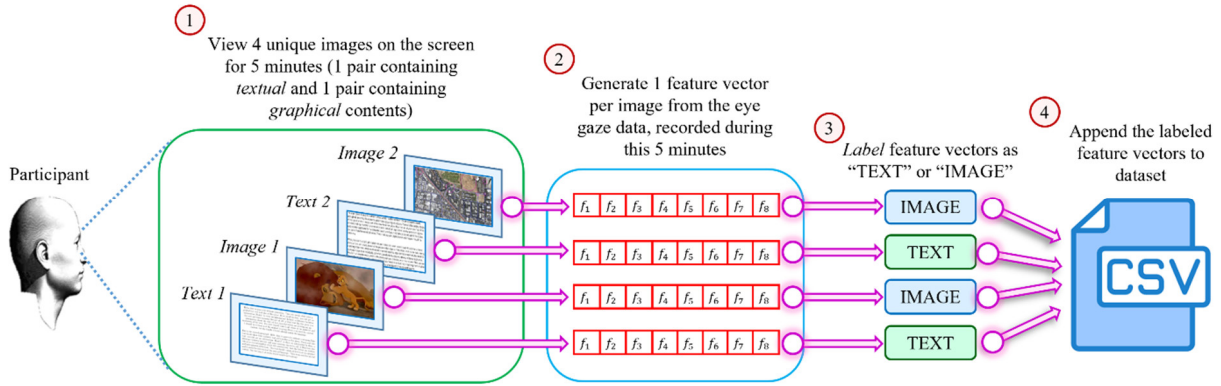

Figure 4: Workflow diagram of generating the training dataset from eye gaze.

Table I: Summary of the dataset prepared for the study.

| Property | Value |
| --- | --- |
| Total Instances | 124 |
| Number of Features | 8 |
| Class Labels | "TEXT" |
|  | "IMAGE" |
| Class Distribution | "TEXT" – 48.39% |
|  | "IMAGE" – 51.61% |

Table II: Extracted features from 10 samples of our dataset.

| Features |  |  |  |  |  |  |  | Label |
| --- | --- | --- | --- | --- | --- | --- | --- | --- |
| *MAX_FC_R* | *MAX_FC_L* | *MIN_FC_R* | *MIN_FC_L* | *AVG_FC_R* | *AVG_FC_L* | *MR_R* | *MR_L* |  |
| 37 | 45 | 2 | 1 | 3.115 | 2.272 | 2.385 | 2.04 | TEXT |
| 363 | 392 | 7 | 3 | 11.274 | 8.318 | 0.227 | 0.374 | IMAGE |
| 119 | 113 | 5 | 6 | 23.462 | 26.857 | 1.066 | 0.758 | IMAGE |
| 17 | 30 | 2 | 2 | 2.175 | 2.635 | 1.894 | 2.147 | TEXT |
| 13 | 29 | 2 | 1 | 2.102 | 1.937 | 1.68 | 2.157 | TEXT |
| 478 | 485 | 9 | 12 | 10.23 | 15.69 | 0.221 | 0.361 | IMAGE |
| 111 | 107 | 1 | 1 | 9.781 | 8.271 | 0.993 | 0.887 | IMAGE |
| 413 | 421 | 4 | 4 | 12.974 | 11.883 | 0.653 | 0.652 | IMAGE |
| 221 | 241 | 4 | 2 | 5.456 | 5.287 | 1.112 | 1.037 | TEXT |
| 521 | 518 | 15 | 13 | 18.641 | 14.391 | 0.576 | 0.489 | IMAGE |

### 3.4 Feature Property Analysis

From the *Feature Histogram-plot* (*FH-plot*) in Fig. 5, where for each feature, the x-axis represents the range of values, and the y-axis represents the frequency of the values in a particular range, it is evident that the *Average Fixation Count* (*AVG_FC*) of both *Left* (*L*) and *Right* (*R*) eyes for *graphical* (images) contents (Fig. 5a) is very high (5 – 10) compared to the *textual* ones (0 – 2.5) (Fig. 5b). The reason for such high values is that from images, people tend to retrieve information from isolated points [7], [8], [11], [24]. Therefore, a greater amount of focus on discrete points increases the fixation count for *graphical* information.

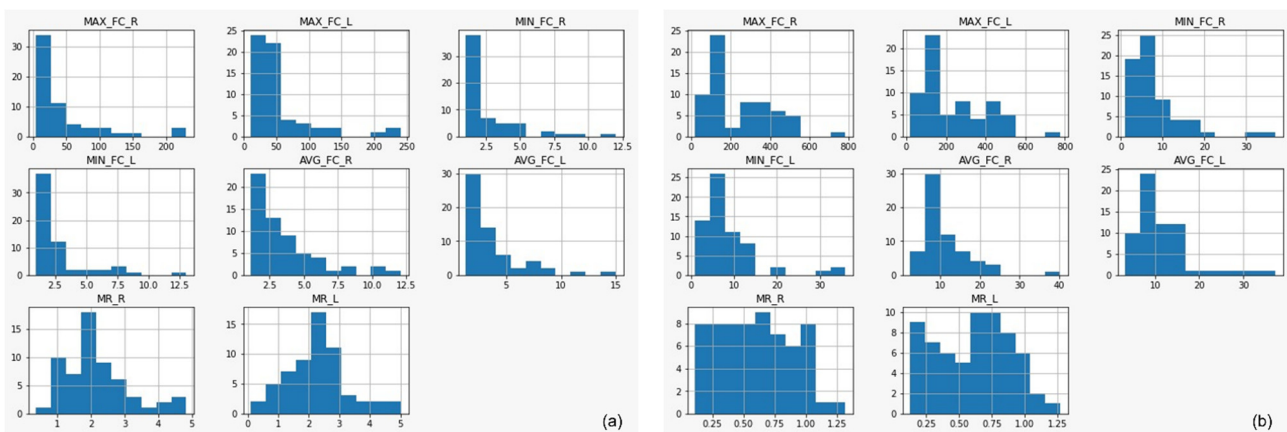

Figure 5: Feature Histogram-plot (FH-plot) of (a) *textual* and (b) *graphical* contents, where for each feature, the x-axis represents the range of values, and the y-axis represents the frequency of the values in a particular range.

Another distinguishable feature is the *Movement Ratio* (*MR*) of the eyes because while retrieving any textual information, a user normally reads from left to right. As a result, the horizontal movement of the eyes dominates the vertical movement [12], [43] and the value of this ratio increases. It can be seen from the *FH-plot* (Fig. 5) that for *textual* contents (Fig. 5a) this ratio is normally distributed within the range of 1 to 5, whereas, for *graphical* contents (Fig. 5b) the value of this ratio ranges between 0 to 1, distributed uniformly. Therefore, a high value of this ratio clearly points to the class of *textual* contents. Based on these observations, *Movement Ratio* (*MR*) of the eyes appear to be the most vital feature, succeeded by *Average Fixation Count* (*AVG_FC*), for classifying visual intention for retrieving *textual* or *graphical* content.

### 3.5 Classification

Our objective is to build a reliable approach of classifying visual intentions based on textual or graphical contents using eye gaze data. Therefore, we adopted a systematic approach of investigating 5 different classifiers such as *K-Neighbors Classifier* (*KNN*), *Gaussian Naïve Bayes* (*NB*), *Logistic Regression* (*LR*), *Support Vectors Machine* (*SVM*), and *Random Forest* (*RF*) to justify which classifier works well with this kind of data. For KNN, we chose the number of neighbors to be 5. We have used *liblinear* solver [44], [45] for *LR, linear* kernel for *SVM*. For *RF*, involving 200 trees with $max\_features = \sqrt{total\_features}$, the quality of split was calculated by Gini impurity [46]. The parameters of the classifiers were optimized using *Grid Search* [47]. Considering the small volume of the generated dataset, the classifiers were trained and tested following supervised learning methodology with *Repeated K-Stratified Fold Cross-Validation* [21] with K=10 and 5 repetitions.

### 3.6 Feature Importance Analysis

In order to perceive the relative importance of the proposed features on the classification of visual intention from eye gaze data, we calculated their Gini impurities [46], using *RF*. For a particular feature, as the mean decrease in this impurity increases, the relative importance for that feature also increases. Utilizing this property, we have generated a *Feature Importance-plot* (*FI-plot*) of our proposed feature vector with the mean decrease in impurity plotted along the y-axis, as shown in Fig. 6, summarizing the relative importance of the features for such classification of visual intention. From the analysis of the *FI-plot*, we have found that the features – eye *Movement Ratio* (*MR*) and *Average Fixation Count* (*AVG_FC*) are two of the most important features, contributing to the classification of visual intention as *textual* or *graphical* using eye gaze data.

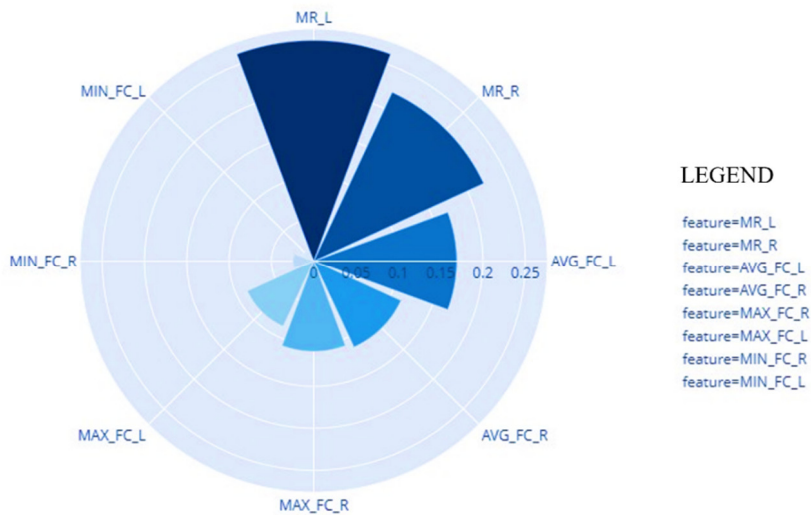

Figure 6: Feature Importance-plot (FI-plot) using Gini impurity, summarizing the relative importance of the proposed features for classifying visual intention for retrieving *textual* or *graphical* information based on eye gaze data. Movement Ratio (MR) of the eyes are the most vital ones, followed by the Average Fixation Count (AVG_FC) for such classification.

### 3.7 Classifier Performance Analysis

After training and testing the classifiers, we have measured their performance using different metrics such as the *Accuracy, Area Under the ROC Curve* (*AUC*), *Precision*, *Recall*, and *F1-score*. From our analysis, we have found *SVM* and *LR* to perform better than the other classifiers.

For any binary classifiers, the *Receiver Operator Characteristic* (*ROC*) curve is a probability curve of the *True Positive Rate* (*TPR*) or "*Sensitivity*", as in (2) and *False Positive Rate* (*FPR*) or "*1 – Specificity*", as in (3). This curve essentially differentiates the signal from the noise. *TPR* and *FPR* quantifies the ability of a binary classifier to correctly classify the *positive* and the *negative* classes, respectively. Therefore, a higher value of *TPR* and a lower value of *FPR* is preferred for any binary classifier to be reliable.

$$TPR \text{ or } Sensitivity = \frac{True\ Positive}{True\ Positive + False\ Negative} \quad (2)$$

$$FPR \text{ or } (1 - Specificity) = \frac{False\ Positive}{True\ Negative + False\ Positive} \quad (3)$$

The *AUC* quantifies the capability of a binary classifier to distinguish between the classes and is used to summarize the *ROC* curve. *Precision* is the ratio of the *True Positive* classifications out of all the positive predictions (*True Positives* and *False Positives*) and *Recall* is a measure of the percentage of actual *positive* classes that were correctly identified. However, maximizing *Precision* may compromise *Recall* and vice versa. To address this issue, the *F1-Score* is used to combine both *Precision* and *Recall* into a single classifier evaluation metric. Finally, *Accuracy* is the ratio of the correct predictions and the total number of predictions available. The evaluation results of each of the classifiers in two scenarios of training and testing (features from both eyes vs one) are summarized in Table III.

Table III: Evaluation of the classifiers using different metrics

| Features Considered | Classifier | Accuracy (%) | AUC | Precision | Recall | F1-Score |
|---|---|---|---|---|---|---|
| Both Eyes | SVM[a] | **92.19** | 0.9548 | **0.9307** | **0.9212** | **0.9206** |
|  | RF | 91.81 | **0.9631** | 0.9282 | 0.9167 | 0.9163 |
|  | LR | 90.71 | 0.9546 | 0.9192 | 0.9069 | 0.9054 |
|  | GNB | 88.92 | 0.9556 | 0.9010 | 0.8890 | 0.8874 |
|  | KNN | 86.15 | 0.9118 | 0.8798 | 0.8605 | 0.8579 |
| One Eye (Left) | SVM | **89.25** | 0.9536 | **0.9139** | 0.8898 | 0.8880 |
|  | RF | 88.92 | 0.9242 | 0.8997 | 0.8893 | 0.8880 |
|  | LR | 89.76 | **0.9552** | 0.9122 | **0.8957** | **0.8945** |
|  | GNB | 88.10 | 0.9533 | 0.8919 | 0.8812 | 0.8793 |
|  | KNN | 85.55 | 0.9248 | 0.8736 | 0.8550 | 0.8518 |

[a] Preferred classifier for the user study in Section 4.

As reported in Table III, the values of the model evaluation metrics are higher when features from both eyes are considered compared to one eye. Thus, we opted for considering eye gaze features of both eyes and *SVM* as our preferred classifier for the user study in the next section, as it has the highest *Accuracy* (92.19%) in this case. The cross-validation *ROC* curve of *SVM*, as shown in Fig. 7, summarizes the *ROCs* (within 1 standard deviation) of all folds using *Repeated K-Stratified Fold Cross-Validation* [21] with K=10 and 5 repetitions and features of both eyes, and the blue line indicates the corresponding mean ROC curve. The average inference time per sample using SVM was about 0.8 milliseconds.

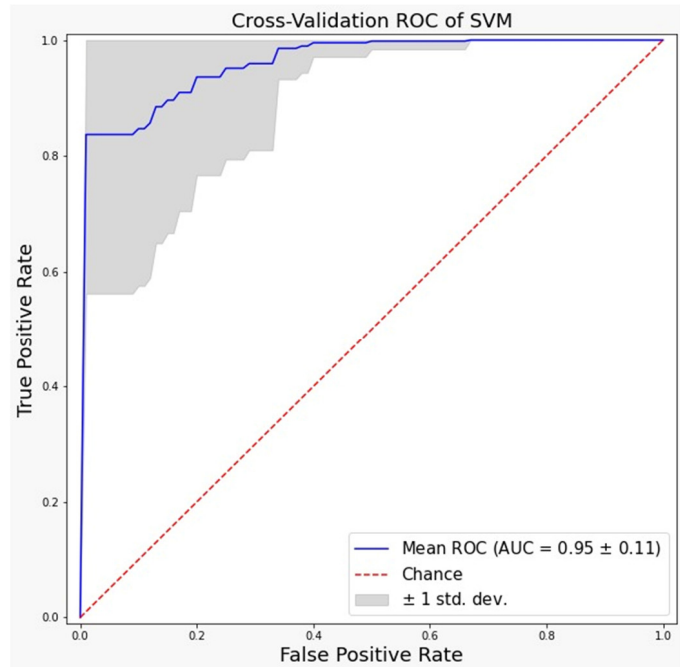

Figure 7: Receiver Operating Characteristic (ROC) curve for SVM.

## 4 USER STUDY: VISUAL INTENTION BY AGE

In this section, we elaborate on the user study where we aim to understand how the interest in *textual* or *graphical* information varied across users of different age groups by analyzing their eye gaze data. An analysis of such behavior will help us understand the design principle behind the development of interactive interfaces, targeted towards users of a particular age group, facilitating enhanced HCI. As mentioned earlier, we adopted the trained *SVM* classifier for this experiment.

### 4.1 Participants

We recruited 30 participants (Mean: 27.17 years, SD: 8.60 years, Male: 56.67%, Female: 43.33%) with informed consent and based on their age, divided them into 3 age groups, where each group was given a tag such as – $G_1$, $G_2$, or $G_3$, 10 participants per group. The demographic details of the 3 age groups are given in Table IV.

Table IV: Demographic details of the 3 age groups considered in the user study.

| Age Group (years) | Group Tag | Mean Age (years) | SD (years) | Male Ratio (%) | Female Ratio (%) |
|---|---|---|---|---|---|
| 16 – 20 | $G_1$ | 17.80 | 1.54 | 70 | 30 |
| 21 – 30 | $G_2$ | 26.60 | 4.10 | 60 | 40 |
| 31 – 45 | $G_3$ | 37.10 | 3.99 | 40 | 60 |

### 4.2 Study Design

After recruitment, each participant was instructed about the 4 tasks that they had to perform. In each of these tasks, they had to go through an interface having *graphical* and *textual* contents. Each participant was given a unique pair of randomly chosen contents per task (*Step 1,* Fig. 8). Each participant was given 6 minutes for each task, during which eye gaze coordinates were recorded, analyzed, and the corresponding feature vector was generated (*Step 2,* Fig. 8). This feature vector was then passed on to the trained *SVM* classifier (*Step 3,* Fig. 8), which determined their visual intention ("TEXT" or "IMAGE") during each task (*Step 4,* Fig. 8). Since, during the experiment, coordinates of the detected pupils were recorded and the corresponding features were extracted, to remove bias in these coordinates, each of these contents was displayed at unique locations on the screen in such a way that they did not overlap. Once the intention was classified, it was tagged with the group tag of the participant (*Step 5*, Fig. 8). The workflow diagram of visual intention detection by age is outlined in Fig. 8. After the visual intentions of all the 30 participants (belonging either to the age group $G_1$, $G_2$, or $G_3$) had been classified as "TEXT" or "IMAGE", for any age group, $G_i$, the *Relative Interest* (*RI*) in *textual* ($RI_{Text}$) or *graphical* ($RI_{Image}$) contents was measured following (4) and (5), respectively.

$$RI_{Text}(G_i) = \frac{Count\ of\ "TEXT"\ as\ inference}{Total\ number\ of\ contents\ for\ group\ G_i} \times 100\% \quad (4)$$

$$RI_{Image}(G_i) = \frac{Count\ of\ "IMAGE"\ as\ inference}{Total\ number\ of\ contents\ for\ group\ G_i} \times 100\% \quad (5)$$

Finally, a semi-structured interview was conducted. For each participant, the experiment lasted for approximately 35 minutes (*4 tasks × 6 minutes per task + 10 minutes interview*). To cross-check the findings of our model, we analyzed the qualitative data, collected during the interview session to find out their actual visual intention during each task. This analysis revealed that the model's detection was within the consensus of the user's interest.

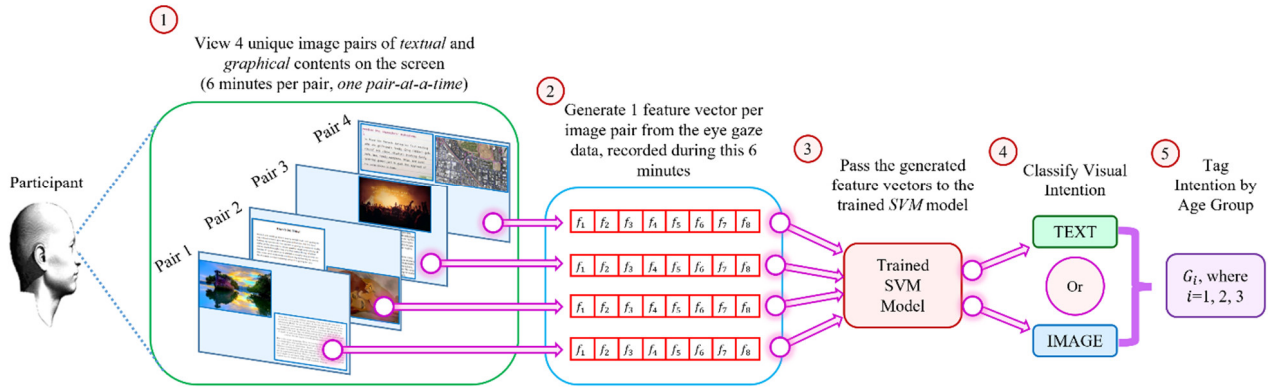

Figure 8: Workflow diagram of visual intention detection by age group from eye gaze.

### 4.3 Result Analysis

Intuitively, visual intention is a subjective cognition process. However, it is of great interest how this behavior varies across users of different age groups. The findings of our experiment provided us with valuable insights into the variance of human visual intention with respect to their age. As seen from Fig. 9, for the young users (Group $G_1$, 16-20 years), the value for $RI_{Image}$, following (5), was found to be 77.50%, meaning that the users within this age group preferred *graphical* over *textual* contents in 77.50% of the cases. Similarly, for the middle-aged users (Group $G_2$, 21-30 years), the values for $RI_{Text}$, following (4), and $RI_{Image}$, following (5), were 42.50% and 57.50%, respectively, maintaining a neutral preference. However, for the elder users (Group $G_3$, 31-45 years), about 70% of them focused on *textual* content.

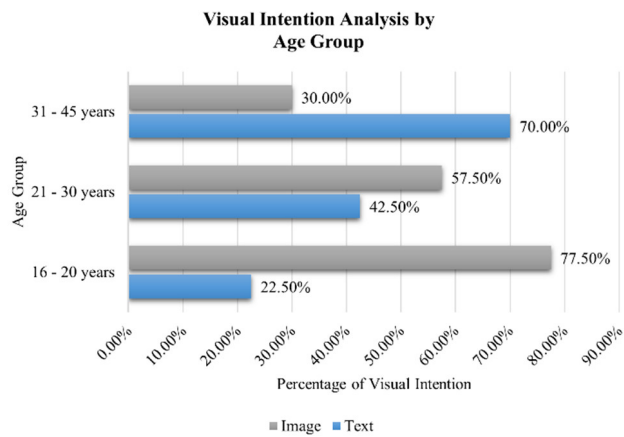

Figure 9: Analysis of visual intention (*textual* or *graphical*) among users of different age group.

From the perspective of design principle, these user preferences, obtained by analyzing eye gaze data, may allow *UI/UX* designers to facilitate the cognitive process of the users of different age groups in HCI. For example, based on our experiment, if we want to design a gender-invariant interface for young users, the proportion of *graphical* contents will have to be higher than the *textual* ones so that maximum exchange of information can be achieved between the user and the interface. Again, to achieve similar goals for middle-aged users, the proportions of these contents will almost be the same due to minimal variation of their preference. For the elderly, however, the proportion of *textual* contents will have to be greater than that of the *graphical* ones.

From the semi-structure interview, we found 29/30 participants were in the consensus with the classifier's detection of intention. Furthermore, we uncovered some applications from participants' responses on leveraging this technique for adaptive user interface design. One area where this can potentially add some interest is online news blogs. We found often participants feel uninterested on the materials due to not having data presented in infographics. However, we also found that the type of content is highly subjective to age. Therefore, having adaptive user interface is likely to increase users' engagement in this area. Based on the idea, if the type of content can be identified as the user's preference, this can also be leveraged for the personal recommendation of contents.

## 5 DISCUSSION

In this work, we have focused on the determination of visual intention to perceive textual or graphical information from interactive interfaces by analyzing eye gaze data in real-time using a low-cost regular webcam. We have tracked and recorded the coordinates of the eye pupils and defined a feature vector containing 8 features that are sufficient for classifying visual intentions for retrieving textual or graphical contents. In this manner, we have analyzed the eye gaze data from 31 users and generated a dataset containing 124 samples, labeled as either TEXT or IMAGE. Using this dataset, we have performed a comparative analysis of 5 different classifiers such as KNN, GNB, RF, LR, and SVM. We have found SVM classifies this type of data reliably, having an *Accuracy* of *92.19%* and an *Average Inferencing Time* of *0.8 milliseconds*. Among the eye gaze features, we have found that eye Movement Ratio (MR) and Average Fixation Count (AVG_FC) are vital for classifying visual intentions as textual and graphical. Furthermore, we have used our trained classifier (SVM) to conduct a user study where we have explored the variation in the relationship of visual intention of a user with respect to age and gender. From this user study, we have observed that the young users prefer graphical over textual contents more than the elder users, with the middle-aged users maintaining a neutral preference between the two contents.

The main motivation of this study was to explore whether a minimal number of eye gaze features can highlight certain user preferences for either textual or graphical contents that may help the UI/UX designers in the process of developing adaptive interactive interfaces, facilitating human cognition. Indeed, from our experimental results, using our proposed features, we have found that analyzing these data can play a vital role in this research area. In this study, we have tracked, recorded, and feature engineered *fixations*, *movement ratios* from eye gaze data using a low-cost regular webcam. We have analyzed the variation of the preference for *textual* or *graphical* contents across users of 3 different age groups and found the preference gradually shifted from *graphical* to *textual* contents with increasing age of the users, as seen from Fig. 9.

In recent times, analyzing user preferences while interacting with a web blog interface [48], [49] has become an emerging area of eye gaze research. The news portal presents similar news in infographics as well as texts. From visual information perspective, information can be presented in line, bar, pie, or tabular format. We can think of multiclass detection of visual intention with different types of information presented using graphs on such portals. Extending to our idea, a potential research direction can be designing an adaptive news portal containing users' preferred infographics or texts based on personalized intention. Therefore, by detecting users' intention from their gaze, we may change the layout of the interface by adapting more graphical or textual contents based on their interest, resulting in an adaptive user interface design.

Another interesting investigation can be identifying users' engagement while visualizing certain data just by using their gaze information. If we can distinguish between graphs viewing patterns by training a classifier with users' gaze information, then we will be able to tell which type of graph is more engaging while giving a summary in a news portal just by analyzing users gaze information during inferencing.

Other potential applications of analyzing eye gaze information can be detecting attention during online classes by analyzing pupil movement patterns, uncovering students' plagiarism behavior from different screens during online proctoring, analyzing gender wise user preferences for certain types of interface color schemes and content types that reduces cognitive load and enhances immersion [50]–[52].

As mentioned earlier, for *textual* or *graphical* contents during training-testing, we have considered aspect ratios of 4:3 and 16:9. Future works should investigate the impact on gaze feature values (i.e., *MR*) having contents with different aspect ratios (e.g., 9:16 and 1:1). Also, the addition of new features such as *pupil diagonal movement variation*, *left to right*, and *right to left movement ratio* are likely to make this system generalizable beyond a fixed aspect ratio. Furthermore, another interesting feature that can be explored is the standard deviation (SD) of the feature AVG_FC.

Potentially, this type of tracker can be used in another avenue – the medical sector. Parkinson's patient encounters eye movement abnormalities. That includes hypometric and slow vertical saccades, normal horizontal saccades, saccadic pursuit. So, if we have a tracker trained on eye features of *Parkinson's patient* vs *Normal person*, then diagnosis of the disease may become easier for the doctor. A similar system can be leveraged for detecting other conditions such as *autism phenotype* [53], *multimodal depression* [54], *medical oculography* [55], *Alzheimer's disease* [56], and *Amyotrophic Lateral Sclerosis* (ALS) [57].

As a concluding remark, our study serves as a proof of concept for certain eye gaze features that can contribute to the design and development of interactive interfaces through the determination of visual intention of users of different age, facilitating their cognitive process while interacting with such interfaces. Subject to further investigation, enhanced user experience through intention aware interactive interfaces, facilitating disease detection can be accomplished through visual intention detection from eye gaze information.

## DECLARATION OF INTEREST

The authors do not declare any conflict of interest that may alter the outcomes of the study in any manner and approve this version of the manuscript for publication.

## ACKNOWLEDGEMENTS

The authors express their heartfelt gratitude to the participants for their valuable time and effort for making this study possible.